\documentstyle[psfig]{l-aa}
\def\ltsima{$\; \buildrel < \over \sim \;$}
\def\simlt{\lower.5ex\hbox{\ltsima}}
\def\gtsima{$\; \buildrel > \over \sim \;$}
\def\simgt{\lower.5ex\hbox{\gtsima}}
\def\gsimeq
{\hbox{\raise0.5ex\hbox{$>\lower1.06ex\hbox{$\kern-1.07em{\sim}$}$}}}
\def\lsimeq
{\hbox{\raise0.5ex\hbox{$<\lower1.06ex\hbox{$\kern-1.07em{\sim}$}$}}}
\def\pn{\par\noindent}

\begin{document}

  \thesaurus{03(11.11.1;       % Galaxies: kinematics and dynamics
            11.19.2;          % Galaxies: spiral
            11.19.6)}         % Galaxies: structure

%03(13.25.2; 11.19.1; 11.09.1: Mkn3)}

  \title{Diffuse Thermal Emission from Very Hot Gas in Starburst Galaxies}

   \author{
	M. Cappi\inst{1}, 
	M. Persic\inst{2}, 
	L. Bassani\inst{1}, 
	A. Franceschini\inst{3}, 
	L.K. Hunt\inst{4}, 
	S. Molendi\inst{5}, 
	E. Palazzi\inst{1}, 
	G.G.C. Palumbo\inst{6,1}, 
	Y. Rephaeli\inst{7}, 
        and P. Salucci\inst{8}}

   \institute{
ITeSRE/CNR, via Gobetti 101, 40129 Bologna, Italy               \and
Trieste Astronomical Observatory, via G.B.Tiepolo 11, 
        34131 Trieste, Italy                                    \and 
Astronomy Dept., University of Padova, vicolo dell'Osservatorio 5, 
        35122 Padova, Italy                                     \and
CAISMI/CNR, Largo E.Fermi 5, 50125 Firenze, Italy               \and 
IFCTR/CNR, via Bassini 15, 20133, Milano, Italy			\and  
Astronomy Dept., University of Bologna, via Zamboni 33, 
        40126 Bologna, Italy                                    \and  
School of Physics and Astronomy, Tel Aviv University, Tel Aviv 69978,
        Israel							\and
SISSA/ISAS, via Beirut 2-4, 34013 Trieste, Italy}

   \offprints{M. Cappi (mcappi@tesre.bo.cnr.it)}

\date{Received..................; accepted...................}

%\titlerunning{{\it BeppoSAX} Observations of Starburst Galaxies}
%\authorrunning{M. Cappi et al.}
   \maketitle

\markboth{Cappi et al.: {\it BeppoSAX} Observations of NGC 253 and M82}{}

\begin{abstract} 
$ $
\pn

{\it BeppoSAX} observations of the nearby archetypical starburst galaxies
NGC253 and M82 are presented. Spectral analysis shows that the 2--10 keV
spectra of both galaxies, extracted from the central $4^\prime$ regions,
are best fitted by a thermal emission model with $kT$ $\sim$ 6--9 keV and
metal abundances $\sim$ 0.1--0.3 solar. The spatial analysis yields clear
evidence that this emission is extended in NGC 253, and possibly also in
M82. These results clearly rule out a LLAGN as the {\it main} origin of
the X-ray emission in NGC 253. For M82, the presence of an Fe-K line
at $\sim$ 6.7 keV, and the convex profile of its 2--10 keV continuum,
indicate a significant thermal component. Contributions from point sources
(e.g. X-ray binaries, supernova remnants, and/or a LLAGN) and Compton
emission are also likely. 
Altogether, {\it BeppoSAX} results provide compelling evidence
for the existence of a hot interstellar plasma in both galaxies, possibly
in the form of superwind outflows from the disks of these galaxies.  
Order-of-magnitude estimates and some implications, such as the expelled
mass and the energetics of the outflowing gas of this superwind scenario,
are discussed.  These new results also suggest some similarity
between the X-ray emission from these galaxies and that from the Galactic
Ridge.

\end{abstract}

\section{Introduction} 

Starburst galaxies (SBGs) are galaxies in which star formation occurrs at
a rate significantly higher than a typical galactic rate. The main
interest for studying SBGs in the local universe stems from the belief
that the starburst phase must have been very frequent during the early ($z
\gsimeq 1$) evolution of galaxies (Madau et al. 1996; Heckman 1999).  
Thus understanding the properties of local SBGs would help in the
characterization of galaxies in the early universe.  Studies of SBGs can
also help to shed light on key astrophysical issues such as:  {\it i)} the
chemical enrichment of the intergalactic medium (IGM) by metal-enriched
starburst-driven outflows (Heckman 1998), {\it ii)} the formation,
evolution and structure of galaxies through the feedback effect provided
by massive stars in the forms of radiation, mechanical energy, cosmic rays
and metal-rich material (Ikeuchi \& Norman 1991); and {\it iii)} the
AGN/starburst connection, since both phenomena require concentration of
material in the nuclear regions and are possibly linked to merging
phenomena and/or the presence of molecular bars (Maiolino et al. 1998).

Evidence for complex galactic-scale outflows driven by starburst activity
has been gathered in recent years. It is primarily based on optical and
low-energy X-ray ($\sim$ 0.1--3 keV: hereafter LEX) observations (Fabbiano
1989 and references therein). These outflows have sometimes been called
``superbubbles'' or ``superwinds'', the latter referring to those
manifestations where the extended hot gas, that emits optical emission
lines and LEX emission, appears to be ejected into the IGM. Optical
emission usually suffers from extinction and is often almost completely
reprocessed into infrared (IR) radiation: therefore the X-ray data, when
available, provide the most direct view of the hot wind material. In
general, results of spectroscopic X-ray studies confirm or, at least, are
consistent with thermal emission from a hot plasma most likely
shock-heated by supernovae (e.g. Dahlem, Weaver \& Heckman 1998: hereafter
DWH98). However, the detailed physical characteristics of the gas --in
particular its chemical abundances-- have remained unclear because the
analysis in the LEX band is complicated by the unknown line-of-sight
extinction, the large uncertainties in the theoretical models used (in
particular around the FeL-shell energy band), and the presence of multiple
temperature components (typically with 0.2 keV $\lsimeq kT \lsimeq$ 3
keV).

At higher energies, available imaging information to date was essentially
limited to the {\it EXOSAT} observation of M82 (Schaaf et al. 1989) and to
the {\it ASCA} observations of a few X-ray bright SBGs [NGC 253, M82: Ptak
et al. 1997 (hereafter P97), Moran \& Lehnert 1997, DWH98; NGC 3310, NGC
3690: Zezas et al. 1998; NGC 1569, NGC4449, NGC2146: Della Ceca et al.
1996, 1997, 1999]. Schaaf et al. (1989) found that the $\sim$1--10 keV
spectrum of M82 was equally well fit with a $\Gamma \sim 2$ power-law
spectrum or with thermal emission characterized by $kT \sim 9$ keV. {\it
ASCA} was able to resolve the 2--10 keV emission from NGC 253 (P97)  but
not that from M82 (Tsuru et al. 1997). Adequate spatial analysis was not
possible even in the case of NGC253. Analysis of the {\it ASCA} spectra
shows that the hard X-ray (hereafter HEX) components of both sources can
be described equally well by either a thermal ($kT \sim 6$--9 keV) or a
power-law ($\Gamma \sim$ 1.8--2.0) model. The puzzling absence, in the
previous data, of significant Fe-K line emission induced several authors
to favour the power-law model: this was interpreted in the framework of
either a low-luminosity active galactic nucleus (LLAGN) (P97; Tsuru et al.
1997; Matsumoto \& Tsuru 1999) or of Compton scattering of relativistic
electrons by the intense FIR ($L_{\rm FIR} \sim 3 \times 10^{10}
L_{\odot}$, see Telesco \& Harper 1980) radiation field (Rephaeli et al.
1991; Rephaeli \& Goldshmidt 1995; Moran \& Lehnert 1997).

In a previous paper (Persic et al. 1998), we have presented the {\it
BeppoSAX} spectral results for the central region of NGC 253, which have
given the first evidence of the 6.7 keV Fe-K line emission and of the
high-energy rollover expected as the signature of thermal emission. In
this paper, we present the results obtained from the temporal (Sect. 3),
spatial (Sect. 4), and spectral (Sect. 5, 6 and 7) analysis of the {\it
BeppoSAX} data for both NGC 253 and M82. As will be shown, our results
suggest a predominantly thermal origin of the HEX component in both
sources. The possible origin(s) and implications of such emission are then
briefly discussed (Sect. 8). In what follows, we assume a distance of 4.8
Mpc and 3.9 Mpc to NGC 253 and to M82, respectively.

\section{Observations and data reduction} 

{\it BeppoSAX} consists of a low-energy concentrator spectrometer (LECS:
Parmar et al. 1997), a medium-energy concentrator spectrometer (MECS:
Boella et al. 1997), a high-pressure-gas scintillation proportional
counter (HPGSPC: Manzo et al. 1997), and a phoswich detector system (PDS:
Frontera et al. 1997).

{\it BeppoSAX} observed NGC 253 on Nov.29 -- Dec.02, 1996 and M82 on Dec.
6-7, 1997. The HPGSPC data will not be considered in the present paper
since the sources are too faint for a correct background subtraction. Our
analysis is restricted to the 0.1--4.5 keV and 1.5--10 keV energy bands
for the LECS and MECS, respectively, where the latest released (September
1997) response matrices are best calibrated (Fiore, Guainazzi \& Grandi
1998). Table 1 reports the log of the observations, with the corresponding
exposure times and derived net (background-subtracted)  count-rates. The
region sizes (Table 1) are chosen in order to minimize the effects of the
extended emission in NGC 253 and to maximize the signal-to-noise ratio in
M82 (but see also Sect. 4). Standard blank-sky files provided by the {\it
BeppoSAX} Science Data Center were used for background subtraction, and
contributed about 10\% (NGC 253)  and 7\% (M 82) of the total counts at 6
keV.

The PDS data reduction was performed using the XAS software package
(v.2.0: Chiappetti \& Dal Fiume 1997). The PDS nominal spectral band is
13--300 keV. The detection of both sources is {\it statistically}
significant (see Table 1).  However, when account is taken of the $\sim$
0.02 cts/s systematic uncertainty, we find that only for M82 is the
detection formally significant (at $\sim$ 4$\sigma$), and only up to
$\sim$30 keV. In the case of NGC 253, no contaminating sources were found
either in the PDS field of view (FOV; which has a triangular response with
FWHM of $\sim$ 1.3$^{\circ}$, see Frontera et al. 1997), nor in the
background regions pointed at $\pm$ 3.5 degrees from the source position
(following the so-called rocking collimator mode). Also in the case of
M82, no contaminating sources were found in the background FOV. However,
the LINER galaxy M81 is only $\sim$ 37$^\prime$ away from M82 so that,
given the PDS triangular response, the measured emission includes half of
the M81 flux. A {\it BeppoSAX} observation of M81 (Pellegrini et al. 1999)
performed in June 1998 indicates that the source was then well fit by a
steep ($\Gamma$ $\simeq$ 1.8 $\pm$ 0.1) power-law spectrum from 0.1 to 100
keV and had a 10--50 keV flux of $\sim$ 4 $\times$ 10$^{-11}$ erg
cm$^{-2}$ s$^{-1}$.  Since the 10--50 keV flux of M82 is 1.7 $\times$
10$^{-11}$ erg cm$^{-2}$ s$^{-1}$, it seems that most of the E$>$10 keV
flux observed from the M82 FOV originates in M81.  The spectrum of M82,
obtained by fitting the PDS data only is, however, much steeper ($\Gamma$
$\simeq$ 3.8 $\pm$ 1.4) than that of M81 which implies only marginal or no
contamination at all. This, combined with the fact that the X-ray flux of
M81 is known to have varied by at least a factor of 4 on a timescale of
months/years (Pellegrini et al. 1999;  Iyomoto 1999), prevents us from
claiming the PDS detection of M82\footnote{In this respect, it is
interesting to note that {\it BATSE} (Malizia et al. 1999) has detected an
even larger HEX (20--100 keV) flux ($\sim$ 5.9 $\times$ 10$^{-11}$
ergs cm$^2$ s$^{-1}$). All this body of evidence suggest that either M81
or M82 do vary significantly at high energies.}. A conservative approach
is thus to consider the M82 PDS data as upper limits, as will be done in
the rest of this paper.

\begin{table}[hbt]
\begin{center}   
{\bf Table 1:} Exposure times and count-rates \\
\begin{tabular}{ccccc}
\hline
Source &  Inst. & Exposure 	& Region size 	& Count-rate \\
       & 	& (ksec) 		& ($^\prime$)	& (10$^{-2}$ cts/s) \\
\hline
NGC 253 & LECS & 55 		& 4		& 3.9$\pm$0.1 \\
        & MECS & 113 		& 4		& 8.9$\pm$0.1 \\
        & PDS$^a$  & 51	& 	78	
& 5.3$\pm$1.7$^{(\sim 3 \sigma)^{b}}_{(\sim 2 \sigma)^{c}}$\\
& & & \\
M82     & LECS & 29 		& 6		& 19$\pm$0.3 \\
        & MECS & 85 		& 6		& 37$\pm$0.2 \\
        & PDS$^a$ & 30 	&	78	
& 7.6$\pm$1.4$^{(\sim 5 \sigma)^{b}}_{(\sim 4 \sigma)^{c}}$\\
\hline
\end{tabular}
\end{center}
$^a$ Energy range considered between 13--50 keV.
\pn
$^b$ Detection significance: statistical errors only.
\pn
$^c$ Detection significance: statistical errors plus systematic 
errors (see text). 
\end{table}

\section{Temporal analysis}

Fig.1 shows, for both NGC 253 and M82, the LECS (0.1--2 keV) and MECS (2--10
keV) lightcurves as well as the LECS/MECS flux-ratio curve. No variability
in either energy band is found for NGC 253.  Using the best-fit model
obtained in Sect. 5, the count-rates measured from NGC 253 correspond to
$F_{\rm 0.1-2 keV}$ $\sim$ 3 $\times$ 10$^{-12}$ erg cm$^{-2}$s$^{-1}$ and
$F_{\rm 2-10 keV}$ $\sim$ 5 $\times$ 10$^{-12}$ erg cm$^{-2}$s$^{-1}$.
These values are consistent with those obtained with {\it ASCA} (P97).

In M82, some short-term ($\sim$ hrs) variability (with $\sim$ 30\%
amplitude) was detected in the MECS energy band at $\gsimeq$ 95\%
significance. The roughly constant hardness-ratio curve (Fig.1, bottom
panel) indicates, however, that the flux variations were not accompanied
by strong spectral variations. To estimate the fraction of the
variable flux in M82, we performed a maximum likelihood fit (see e.g.
Maccacaro et al. 1988) of the count-rate distribution of M82, and obtained
a dispersion of the light curve around its mean value of only 15\% $\pm$
4\%. In conclusion, the lack of spectral variability and small dispersion
around the mean suggest that this variable component should not affect
significantly the spectral analysis and related inferences presented
in this paper.

The short-term variations in M82 observed with {\it BeppoSAX} are similar
to those reported by Matsumoto \& Tsuru (1999) and Ptak \& Griffiths
(1999) from {\it ASCA}. These authors also report long-term HEX (2--10
keV) variations by up to a factor of $\sim$ 4 on a time scale of
several days, suggesting the presence of a LLAGN in M82 (see Sect. 8).  
Using the best-fit model presented in Sect. 5, the observed count-rates
correspond to $F_{\rm 0.1-2 keV}$ $\sim$ 1.1 $\times$ 10$^{-11}$ erg
cm$^{-2}$s$^{-1}$ and $F_{\rm 2-10 keV}$ $\sim$ 2.9 $\times$ 10$^{-11}$
erg cm$^{-2}$s$^{-1}$, respectively. The low-energy flux is consistent
with the (constant) value obtained by {\it ASCA}; the flux in the higher
energy band is about equal to the average of the values measured with {\it
ASCA} (Ptak \& Griffiths 1999), implying that M82 was in a normal/typical
state during the {\it BeppoSAX} observation.

\section{Spatial analysis} 

The LECS relatively low spatial resolution (FWHM $\sim 10^{\prime}$ at 0.3
keV)  and low statistics do not allow a detailed spatial analysis at
$E~\lsimeq~2$ keV.  Higher resolution images at low energies were
previously obtained with {\it Einstein}, {\it ROSAT} and {\it ASCA}
(Fabbiano \& Trinchieri 1984; Read, Ponman \& Strickland 1997; DWH98; P97, 
and ref. therein), and show much finer details.

Spatially, {\it BeppoSAX} can do better than any previous observation at
E$>$2 keV because of the unprecedented MECS statistics and spatial
resolution (at 6 keV, the radii --measured from the center of the FOV--
encompassing 50\% and 80\% of the observed flux are $r_{50\%}$ $\sim$
1$^\prime$ and $r_{80\%}$ $\sim$ 2$^\prime$, i.e. about twice better than
{\it ASCA}). The analysis presented in this section focuses on the data
above 3 keV which are dominated by the HEX component (see Sect. 5). Fig.2
shows the MECS 3--10 keV images of both galaxies superimposed on Digital
Sky Survey data. {\it a)} The left panel of Fig.2 clearly shows one
of the most interesting results of the present analysis: the HEX emission
of NGC 253 is extended and elongated along the optical major axis. In NGC
253, there is also an indication for an X-ray ``cone'' extending
southwest.  No point sources embedded in the extended emission are
detected, but within the MECS resolution their presence cannot be ruled
out firmly. Another interesting feature in NGC 253 is the apparently
extended emission along the minor axis (northwest to southeast),
reminiscent of the ROSAT PSPC and HRI results (R97; DWH98). Indeed, such
morphology could be the signature of very hot gas ejected out of the
galaxy (see Sect. 8). {\it b)} The right panel of Fig.2 suggests the
presence of a more symmetric X-ray halo in M82, although with some 
apparent excess emission (at a $\sim$2$\sigma$ level) from the outer
galactic regions along the optical minor axis in the northwest direction.
The shift between the X-ray and the optical centroids is within the
systematics ($\sim$ 1$^\prime$) of the absolute position reconstruction
for {\it BeppoSAX}.

The radial profiles of the 3--10 keV emission from NGC 253 and M82 are
shown in Fig.3, together with the instrumental PSF (energy-weighted over
the source spectra). The HEX emission in NGC 253 extends up to $\sim$
8$^{\prime}$ ($\simeq$ 11 kpc). There is also marginal evidence that
the emission from M82 extends to $\sim$ 5$^{\prime}$ ($\simeq$ 6
kpc). If real, the extended component of M82 contributes $\sim$ 10\% of
the total 3--10 keV flux. (A more quantitative assessment of the reality
of this spatial extent as evidenced in the {\it BeppoSAX} data is not
possible because of PSF and background subtraction systematic uncertainties, which
are crucial at these levels of precision.)

In NGC 253, the extension is also evident if one considers the 3--5 keV,
6--7 keV and 7--10 keV datasets separately (Cappi et al. 1999), indicating
Fe-K-line emission region is extended.  At all radii, the 3--10 keV
surface brightness profile can be fitted with a King model ($S=S_{0} [
1+(r/r_{c})^{2}]^{-\alpha}$, where $S$ is the surface brightness, $r$ is
the distance from the center, $r_{c}$ is the core radius, and $\alpha$ is
an index) having $r_{c}$ $\sim 1.5^{\prime}$ ($\sim 2$ kpc) and $\alpha$
$\sim$ 1.3. For $r$ $\geq 2^{\prime}$, the brightness can be fitted with a
simple $r^{-2}$ profile consistent with the emissivity expected from an
expanding gas dominated by ram pressure (see Sect. 8.2.6). We find that
the diffuse component in NGC 253 contributes at least 40\% of the total
3--10 keV flux within $\sim$ 8$^{\prime}$. If the extension in M82 is
real, then the diffuse component is $\sim$ 10\% of the total 3--10 keV
flux.

\section{Spectral analysis}

For the spectral analysis, the LECS and MECS data were rebinned so as to
sample the instrumental resolution with 3 and 5 channels per energy bin,
respectively (corresponding to at least 20 and 50 cts bin$^{-1}$). The PDS
data were grouped logarithmically between 13 and 60 keV in order to
achieve a compromise between the highest number of bins and the highest
S/N ratios.  Given the large uncertainties on the PDS data points (a low
S/N for NGC 253 and the possible contamination of the M82 emission by
M81), we did not include them in the fitting procedures (but we show them
in the figures). Galactic column densities of N$_H=1.28$ $\times$
10$^{20}$ cm$^{-2}$ for NGC 253, and N$_H=4.27$ $\times$ 10$^{20}$
cm$^{-2}$ for M82, were added in all the spectral fits (Dickey \& Lockman
1990). The spectral analysis was performed using version 10.00 of the
XSPEC program (Arnaud 1996).

\subsection {Continuum}

Fig.4 shows the results obtained by fitting the source spectra with a
single power-law model. Consistent with results from previous satellites
(see P97, and refs. therein), the $E<2$ keV data of both sources clearly
require at least one thermal component. At $E>2$ keV, a pure power-law fit
to the continuum, which was allowed by the earlier {\it ASCA} data (P97),
seems to be ruled out by the present {\it BeppoSAX} data (owing to the
unprecedented statistics at $E\gsimeq 5$ keV), for both NGC 253 (see
Persic et al. 1998) and M82. In the case of M82, fitting the LECS+MECS
2--10 keV data by a model that involves a bremsstrahlung (power-law)
continuum plus 3 gaussian emission lines gives $\chi^2=$ 78 (92) for 91
d.o.f.: thus, a power-law instead of thermal model is significantly less
likely ($\Delta \chi^2$=14).  \footnote{ When the PDS data --if considered
as actual detections-- and/or the $E<2$ keV data are also taken into
account, then the power-law model seems even less likely when compared
with the thermal model (typically, $\Delta \chi^2 \gsimeq 20$). Matsumoto
\& Tsuru (1999) likewise note that the HEX component detected with {\it
ASCA} in M82 is statistically better described by a thermal model than by
a power-law component.} The situation is similar for NGC 253. These
results can also be deduced from the behaviour of the $\gsimeq$5 keV MECS
and PDS upper-limit residuals (see Fig.4). Even ignoring the Fe-K
line detection (see below), these residuals show a clear steepening of the
continuum toward higher energies, which may by itself be considered
as evidence for thermal emission. Whatever the model used for the HEX
continuum (i.e a power-law or a thermal model), it needs to be absorbed in
order not to over-produce the continuum at E $\lsimeq$1 keV (as was also
deduced from {\it ASCA} data: see P97).

We then fitted the spectra with a two-temperature bremsstrahlung model
(the HEX component being absorbed) plus gaussian lines in order to define
the properties (energy and intensity) of the detected emission lines (see
Fig.5 and Table 2). At energies lower than $\sim$ 3 keV, several lines are
present in the spectra (e.g. from FeL and H- and He-like O, Ne, Mg, Si and
S ions): these lines are typical of LEX thermal components (see also the
{\it ASCA} results of P97). One of the most remarkable results of the
present analysis is that an Fe-K line at E $\sim$ 6.7 keV is clearly
detected in both galaxies, confirming previous detections from the same
data (Persic et al. 1998 for NGC 253), and from {\it ASCA} data (Ptak et
al. 1997 and Matsumoto \& Tsuru 1999 for M82).

To model the broad-band spectra of both galaxies, we then tried several
thermal models:  pure bremsstrahlung, single-temperature Raymond-Smith
(RS), multi-temperature RS, Mewe-Kaastra-Liedahl ($mekal$; Mewe et al.
1985), and combinations of them, with and without absorption, and with
solar and variable abundances. We found that the best description of both
galaxies' spectra is in terms of a two-temperature $mekal$ model with
variable abundances ($vmekal$ in XSPEC), as shown in Table 3 and Fig.6. We
preferred the $mekal$ to the RS model (in spite of the equally good fit
obtained) because of the more recent emissivities (notably for Fe) used in
the former (Liedahl et al. 1995). Solar abundance ratios from Anders \&
Grevesse (1989) are used throughout.

\subsection{Emission lines}

In order to reduce the number of free parameters (following Persic et al.  
1998), the abundances of He, C, and N were set equal to the solar values,
while the heavier elements were divided into two groups: Fe and Ni, which
are most likely associated with SNe I products, and $\alpha$-elements O,
Ne, Mg, Si, S, Ar and Ca, likely associated with SNe II products.

Elements in the same group were constrained to have the same abundances in
solar units. Furthermore, the abundances of the $\alpha$-elements related
to the HEX component were set equal to those of the LEX component after
verifying their mutual consistency (within errors) when left free to vary.

The detection of the Fe-K line allows us to reliably determine the Fe
abundances of the line-emitting material: we determine a value of about
0.3 and 0.07 solar for NGC 253 and M82, respectively. The former value is
consistent with the upper limits derived from previous X-ray observations.
The latter, unusually low value may hint at the presence of an appreciable
non-thermal contribution that would raise the local continuum and swamp
the iron line (see Sect. 7). Similarly, low Fe abundances are found for
the LEX component, but current uncertainties of theoretical models in
computing the Fe-L emissivity (e.g. Liedahl et al. 1995) makes this result
less certain.  On the other hand, the $\alpha$-elements seem to have
larger abundances ($\sim$ 1.6 solar for NGC 253 and $\sim$ 2.9 solar for
M82). However, we caution again that these numbers are highly uncertain
because of the very likely multi-temperature (and possibly
non-equilibrium) nature of the emission (as found in other SBGs, see Della
Ceca et al. 1997 and DWH98). Assuming such multi-temperature plasma models
would likely result in very different abundances for the $\alpha$-elements
(e.g. DWH98). Moreover, given the complexity of the models and the poor
statistics of the data at low energies, a significant degeneracy between
derived abundances and fitted column density is unavoidable. In view of
these considerations, the $\alpha$-element abundances derived here should
be considered as purely indicative.

\section{Combined spectral \& spatial analysis: NGC 253}

The results obtained from the spatially-resolved spectral analysis of NGC
253 are shown in Fig.7. The analysis refers only to the MECS data since
the quality of the LECS data statistics does not allow a more detailed
study.

Spectral fits have been performed in the 3--10 keV band in each of 9
concentric annuli centered on the source and having similar statistics.
Based on these fits it is concluded, at a significance level $>$ 99.95\%,
that the temperature decreases with increasing distance from the center.
Similar results were obtained from spectral fitting of three different
circular regions: center, north and south. These fits show that: {\it (a)}
the Fe-K line is detected in all three regions (at $>$ 99.99 \%, 90\% and
68\%, in the center, north and south, respectively); and {\it (b)} the
temperature at the center, $kT$ $\sim$ 6.0 $\pm$ 0.4 keV, is significantly
higher than in both the north, $\sim$ 3.7 $\pm$ 0.5 keV, and south, 3.7
$\pm$ 0.6 keV, regions. We are not aware of any instrumental effect that
could produce such large differences (despite the poor statistics
available) since the {\it BeppoSAX} PSF does not vary significantly as a
function of energy within the (limited) off-axis distances considered
here. No significant spatial variations of either the absorption or the Fe
abundance are found.  One should, however, bear in mind the limited energy
range (1.5--10 keV) available from using only the MECS data for
determining the absorption and the limited statistics (and
correspondingly large errors) which do not allow a definite measurement of
the iron abundance at large radii.

A similar analysis in the case of M82 was not possible because of its
smaller, and only marginally significant, spatial extension. It
should be noted, however, that analysis of {\it ASCA} data (Ptak \&
Griffith 1999; Matsumoto \& Tsuru 1999) presumably shows that X-ray
emission comes mostly from the central $10^{\prime \prime}$ region of
M82.

\section{Model-dependence of the best-fit results}

Given the various possible contributions to the emission from SBGs, it is
essential to quantify the effect of adding an extra power-law component to
the best-fit two-temperature model discussed above (Sect. 5). Such an
additional spectral component may be due to nonthermal emission from a
LLAGN, X-ray binaries (XRBs), supernovae remnants (SNRs), as well as
Compton scattering (e.g., Rephaeli et al. 1991). In particular, in
M82 some short-term ($\sim$ hrs) variability (with $\sim$ 30\%
peak-to-valley variation) is seen in the light curve (see Sect. 3),
possibly indicating a contribution from an AGN to the HEX flux. Given the
lack of strong point-sources in the {\it ROSAT} PSPC/HRI observations of
M82 (DWH98; Bregman et al. 1995), these HEX sources could be either
highly-variable (see P97) or strongly absorbed at $\leq$3 keV. An
appreciable non-thermal HEX contribution would presumably result in a
reduction of the gas temperature obtained from our best fit, and
consequently in an increase of the inferred metal abundances in M82.

We now consider the effect of a non-thermal component in the
framework of our derived best-fit model.  We repeat the data analysis
(with the PDS upper limits included)  using a photon index in the range $1
\leq \Gamma \leq 2$ and several values of photoelectric absorption, N$_H$
$\geq$ 10$^{22}$ cm$^{-2}$. We find that a power-law contribution of
$\lsimeq$ 20\% and $\lsimeq$ 10\% (in the 3--10 keV range) for NGC 253 and
M82, respectively, does not improve the fit. In NGC 253, by forcing a
$\Gamma$=2 power-law contribution to be 50\% of the total 3--10 keV flux,
the plasma temperature decreases from $kT$ $\sim$5.7 keV to $\sim$4.4 keV
while the iron abundance increases from $\sim$0.3 to $\sim$0.5 solar: the
fit, however, is slightly worse ($\Delta \chi^2$ = 5). Likewise, in M82 the
plasma temperature decreases from $kT$ $\sim$8 keV to $\sim$6 keV, and the
iron abundance increases from about 0.08 to 0.25 solar: however, the fit
is statistically worse ($\Delta \chi^2$ = 11).  \footnote{ Alternatively,
setting the abundances of M82 equal to 0.3 solar, we find that $\gsimeq$
50\% of the 3--10 keV flux must originate from an additional HEX, absorbed
component. Likewise, the fit is statistically worse.} Because the average
XRB spectrum may not be well described by a single absorbed power-law but
could require the addition of an Fe-K line emission (most XRBs are known
to emit relatively strong Fe-K lines at 6.4 keV and, in some cases, at 6.7
keV) the abundance differences between the cases with and without
power-law component may be smaller than what is suggested by this spectral
analysis.

To summarize, some contribution from a non-thermal component, while not
required (though expected on theoretical grounds), is allowed by the {\it
BeppoSAX} data. An added motivation for the existence of a non-thermal
component in M82 might also be the uncomfortably low iron abundance that
is otherwise deduced. The strongest motivation, however, comes from
the temporal and spatial analysis of {\it ASCA} data (Ptak \& Griffiths
1999; Matsumoto \& Tsuru 1999). These studies suggest that a HEX
component, responsible for most of the 2--10 keV flux and for its time
variability, originates from a LLAGN embedded in the nucleus of M82. The
fact that we do not see spectral evidence for a power-law component in the
{\it BeppoSAX} data, may be explained by the relatively low flux state of
M82 during our observations, when the putative LLAGN was possibly
quiescent.

\section{Discussion}

\subsection{Origin of the LEX emission}

The temperature and flux of the low-energy (0.1--2 keV) component for both
sources are similar to those obtained from previous observations (compare
our results in Table 3 with those of {\it ASCA} in Tables 1 and 2 of P97).
This is consistent with the conclusions of various authors (P97, DWH98 and
references therein) that the LEX emission is most likely the result
of both point-source radiation (bright SNRs, X-ray binaries etc.) and of
diffuse emission from hot gas, due to either merged SNRs or material
shock-heated by the starburst-driven winds. Specifically, for both
galaxies, the luminosity of the LEX components is a small fraction
($\lsimeq$ 1\%) of the total power ($\sim 3 \times 10^{42}$ erg s$^{-1}$)
available, assuming a SN rate of $\sim$ 0.1 event yr$^{-1}$ (Antonucci \&
Ulvestad 1988, Wei$\ss$ et al. 1999), plus a typical SN power of $10^{51}$
ergs.

The chemical abundances in the low-energy component, deduced for the iron
and for the $\alpha$-elements Ne, Mg, Si and S, are in agreement with the
predictions of chemical evolution models for SBGs (Bradamante et al. 1998)
and superwind models (Suchkov et al. 1994), and hence they too favour the
SB-driven galactic superwind picture (Heckman et al. 1990).  In such
models the galactic wind is powered mostly by Type II SNe (which are
predominant [Marconi et al. 1994] and, in fact, dominate the energetics
during starbursts) and minimally by Type Ia SNe. Consequently, the
resulting wind is overabundant in $\alpha$-elements (ejected during Type
II SNe explosions) and underabundant in Fe (of which only a small fraction
comes from Type II SNe, whereas the bulk comes from Type Ia SNe). For He,
C and N, which are restored by low- and intermediate-mass stars through
stellar winds, the predicted abundance is essentially solar. It should be
reminded, however, that these considerations should be taken as
order-of-magnitude (see Sect. 5).

Similar thermal X-ray components with temperatures corresponding to $kT
\sim$ 0.5--1 keV, often subsolar abundances, and luminosities in the range
$10^{38}$--$10^{40}$ erg s$^{-1}$ have been observed with {\it Einstein},
{\it ROSAT} and {\it ASCA} in a number of other nearby spirals and
starburst galaxies (e.g.: Fabbiano 1989, DWH98, Iyomoto et al. 1996;
Serlemitsos et al. 1995;  Della Ceca et al. 1996, 1997). These studies
suggest a common underlying ``driver'' for the LEX component, i.e. SNRs,
whose strenght is proportional to the SN rate. As pointed out in Persic et
al. (1998), the important analogy with the Galactic Ridge X-ray Emission
(GRXE) should also be stressed.  {\it BeppoSAX} spectra of NGC 253 and M82
(Fig.6) look indeed remarkably similar to the {\it ASCA} 0.4--10 keV
spectrum of the GRXE (see Fig.3 in Kaneda et al. 1997). Moreover,
rescaling the above energetic arguments to a lower Galactic SN rate of
$\sim$ 0.01 yr$^{-1}$, the GRXE is again $\sim$ 1\% of the total energy
available from SNe suggesting similar efficiency in transforming
mechanical energy into X-ray emission.

In conclusion, the complex LEX X-ray emission from both galaxies is 
consistent with contributions from mostly SNRs plus a SB-driven superwind.

\subsection{Origin of the HEX emission}

The origin of the HEX component in NGC 253 and M82 cannot yet be
determined. Any viable model of its origin should account for the
following {\it BeppoSAX} observational results: thermal-like continuum
characterized by a $kT \sim$ 5--8 keV, 6.7 keV Fe-K emission with EW
$\sim$ 400 eV (NGC 253)  and 60 eV (M82), photoelectric absorption column
density of $\sim$ 10$^{22}$ cm$^{-2}$, $L_{2-10 \rm keV}$ $\sim$ 10$^{40}$
erg s$^{-1}$, spatially extended continuum and Fe-K emission (certainly in
NGC 253 and possibly in M82), and $\sim$ 30\% short-term flux variability
in M82.

As in the case of the low-energy component, the present {\it BeppoSAX}
results highlight an interesting, general similarity with the {\it ASCA}
observations of the GRXE (Kaneda et al. 1997). Also in this case, a
high-energy thermal component is detected with similar temperature and
absorbing column density. If scaled for a $\sim$ 1/10 times lower Galactic
SN rate, the power in the HEX component is comparable to that from the
Galactic ridge. The only spectral difference is an appreciably higher,
$\sim$ 0.8 solar, metal abundance deduced from the GRXE.  As already
recognized by Heckman, Lehnert \& Armus (1993), the X-ray data from the
plane and the central regions of the Milky Way (Koyama et al. 1986;
Yamauchi et al. 1990; Kaneda et al. 1997) are probably one of the best
pieces of evidence for a Galactic-scale superwind. This apparent
similarity suggests a common physical origin. We note that the origin of
the high-energy component from the GRXE (Bleach et al. 1972; Wheaton
1976; Valinia \& Marshall 1998, and ref. therein) is still unknown (e.g.
Tanaka et al. 1999).

As shown in Sect. 7, a contribution from an additional
absorbed high-energy component is consistent with the {\it BeppoSAX}
data. This caveat should be kept in mind when
discussing several hypothesis on the nature of this HEX emission
(see below).

\subsubsection{Supernova Remnants}

As pointed out in Sect. 8.1 for the LEX component, the total energy
available from SNRs in these two starburst nuclei is certainly sufficient
to drive the X-ray emission power. However, assuming a typical $L^{\rm
SNR}_X \sim 2 \times 10^{35}$ erg s$^{-1}$, then a large number of SNRs
($\gsimeq~ 8 \times 10^{4}$ for NGC 253 and $\gsimeq~ 3 \times 10^{5}$ for
M82) are required to power the HEX component while, assuming a SNR shock
front of 50 pc radius and a starburst region of 5 kpc radius and 0.5 kpc
thickness, $\leq 7.5 \times 10^{4}$ SNRs can be at work if the two-phase
state of the interstellar matter (ISM) in NGC 253 and M82 are to be
preserved.  Even neglecting these apparent difficulties, the biggest
problem for the SNRs hypothesis is the spectral discrepancy: SNRs usually
have substantially lower temperatures ($kT~ \lsimeq$ 3--4 keV) and
typically much larger abundances.  Therefore, we rule out SNRs as {\it
major} contributors to the HEX emission.
 
\subsubsection{Compton Scattering}

It has been realized that one of the manifestations of the enhanced star
formation and SN rates in SBGs is a corresponding increase in the efficiency
of the acceleration of protons and electrons to high energies. The main
consequence of a strong energetic proton component is a higher rate of
ionization and heating of ISM, whereas relativistic electrons give rise to
more intense synchrotron and Compton emissions.  Moreover, the strong FIR
emission from warm dust (heated by stellar emission) further intensifies the
emission from Compton scattering of the relativistic electrons on the FIR
radiation field (Schaaf et al. 1989;  Rephaeli et al. 1991). A detailed
convection-diffusion model was proposed for electrons in NGC 253, with the
energy density of the electrons normalized by the observed radio flux and
estimated mean magnetic field (Goldshmidt \& Rephaeli 1995). Calculation of
the predicted high energy X-ray emission from Compton scattering of the
electrons off the measured FIR emission yielded a 50--200 keV flux in good
agreement with what was claimed to have been detected from this galaxy by the
{\it OSSE} experiment aboard {\it CGRO} (Bhattacharya et al. 1994).  
Irrespective of the reality of the {\it OSSE} detection, the model provides
adequate description of the expected Compton emission from NGC 253.

As discussed in Section 2, the 13--50 keV NGC 253 flux measured by the PDS is
only marginally significant, so a detailed comparison between the model
predictions and the PDS data is not meaningful.  It is nonetheless
interesting to compare the PDS flux level in this band - when viewed as an
upper limit - with the level predicted based on the Goldshmidt \& Rephaeli
(1995) model. Integrating the calculated spectrum from the disk and halo over
the band, we compute a flux which is consistent with the PDS upper-limit
within a factor of 2. Given the qualitative nature of this comparison and
uncertainties in the PDS analysis, the only general conclusion that can be
drawn from this rough similarity (in the values of the predicted and measured
fluxes) is that the residual (after subtraction of the thermal emission)  
HEX emission from NGC 253 could indeed be accounted for by Compton
emission.

The similarity of the power-law radio index in M82 and best-fit index deduced
from a fit to the {\it ASCA} measurements, led Moran \& Lehnert (1997)  to
suggest that the main component of the complex emission from this galaxy is
due to Compton scattering, whereas P97 found that the `harder' component is
equally well fit by either thermal or power-law model. The evidence from our
{\it BeppoSAX} measurements of significant thermal emission from hot gas
which also gives rise to the Fe XXV $K_{\alpha}$ emission, weakens the
need to
invoke an appreciable nonthermal component at energies below 10 keV, but the
PDS upper-limits in the 13--50 keV band do allow the presence of considerable
nonthermal emission in M82. An attempt to quantify the level of this emission
in M82 is being made in the ongoing analysis of the results of {\it RXTE} 
measurements of this galaxy (Rephaeli \& Gruber 1999).

\subsubsection{X-ray binaries}

Low-mass and high-mass X-ray binaries (LMXRBs, HMXRBs) may significantly
contribute to the total X-ray emission of SBGs. While this is certainly
true for their LEX emission, there are several problems for the high
energy component. As in the two previous cases, their average X-ray
spectra are substantially different from those observed in NGC 253 and
M82. Maybe an appropriate combination of LMXRBs and HMXRBs (typically
described by exponentially cut-off power-laws with index $0.5 \lsimeq
\Gamma \lsimeq 2.5$) could reproduce a single-temperature thermal
continuum, but few of these sources exhibit 6.7 keV lines (they, instead,
do show Fe-K lines at 6.4 keV).

The 30\% flux variations observed in M82 with {\it BeppoSAX} can be taken at
first glance as evidence for a significant contribution from X-ray binaries
or galactic black-hole candidates (P97). This would require at least a few
very strong sources, with $L_{2-10 keV}$ $\sim$ 10$^{40}$ erg s$^{-1}$. Even
ignoring the theoretical difficulties in accounting for such high-luminosity
stellar sources, there are several considerations pointing against this
scenario; among these: such sources have not been resolved by the ROSAT
HRI/PSPC observations (DWH98; but note that strong low-energy absorption may
help to explain such discrepancy). Another consideration is that the factor
of $\sim$ 4 variability observed by {\it ASCA} would require that several of
these sources have varied in phase, a condition which seems rather unlikely
(Ptak \& Griffiths 1999; Matsumoto \& Tsuru 1999).

Finally, an XRBs scenario for NGC 253 would require two different
populations of XRBs in order to explain the lower temperature at larger
radii: one population with harder spectra near the nuclear region, and
another one with softer spectra in the outer disk regions.

Thus, we conclude that it is unlikely that XRBs are sources of a significant
part of the HEX component.

\subsubsection{Hidden LLAGN}

The substantial photoelectric absorption obtained from our best-fits
require a high hydrogen column density, $N_H \sim 0.6-1.2 \times 10^{22}$
cm$^{-2}$. M82 shows short-term (e.g. this work) and long-term (see Ptak
\& Griffiths 1999) variability. These results suggest that there could be
a considerable contribution from a buried LLAGN, with the dust (abundant
in actively star-forming regions) explaining the observed intrinsic
absorption.  It should be noted that in this hypothesis we do not
differentiate between a LLAGN and a (possible) medium-mass black hole
candidate with $M_{\rm BH}$ $\sim$ 10$^{2-4}$ M$_{\odot}$ (see Ptak \&
Griffiths 1999).

However, the {\it BeppoSAX} spectra of both M82 and NGC 253 strongly prefer a
thermal description for the HEX emission rather than a power-law
component, ubiquitous in AGNs (see Sect. 5). The LLAGN scenario is certainly
ruled out in the case of NGC 253 because the {\it BeppoSAX} data have shown
that both its HEX continuum and its Fe-K line emission are extended,
and that this extended component contributes more than 40\% to the total
3--10 keV emission.

Nevertheless, as detailed below, we cannot exclude some significant, but
not dominant, LLAGN contribution in M82. The {\it BeppoSAX} observations
can be reconciled with the conclusions obtained from the long-term {\it
ASCA} studies (Ptak \& Griffiths 1999; Matsumoto \& Tsuru 1999) if
one supposes that during the {\it BeppoSAX} observation the central LLAGN
was in a relatively low state and its HEX emission was swamped by the
thermal emission associated with the starburst itself.

\subsubsection{A spectral conspiracy?}

Another possibility would be to consider a combination of SNRs with emission
from Compton scattering, XRBs and/or a LLAGN. In this case, SNRs are a
necessary ingredient in order to produce the observed 6.7 keV line and would
contribute with a lower (than observed)  temperature thermal spectrum. The
second component (Compton scattering, XRBs and/or LLAGN) would contribute
with an absorbed power-law component. The combination of the two could then
``mimic'' the measured high-temperature thermal HEX component. A similar
explanation, although with a different physical interpretation, has been
proposed by Valinia \& Marshall (1998) to explain the high-temperature
emission from the GRXE.

Even though not warranted, a ``spectral conspiracy'' may be possible 
within the {\it BeppoSAX} data (see sect.7). It should be
noted, however, that
in such a case it would be more natural to combine SNRs, Compton
scattering and/or XRBs in order to explain the extension of the HEX
emission from NGC 253, while a contribution from a LLAGN would explain the
short- and long-term variability observed in M82. Higher quality data,
such as those expected from the {\it Chandra}, {\it XMM} and $ASTRO-E$
satellites, could possibly clarify the contribution from these different
components. In particular, if our hypothesis for M82 is correct, we
predict that the equivalent width of the Fe-K line (produced by thermal
emission from SNRs) should increase with decreasing 2--10 keV flux (which
is affected by the time-varying LLAGN power-law contribution).

\subsubsection{The Superwind Model} 

In this subsection, we consider a simple, alternative explanation for the
HEX component that requires no spectral fine-tuning. Such an alternative
explanation is that the emission is truly originating from a hot thermal
plasma. Our spatial analysis shows that this plasma is extended, at least
in NGC 253, along the disk and possibly along its minor axis direction
(Sect. 4, Fig.2). In both the HEX and LEX emissions, the diffuse emission
we observe may be related to the same superwind phenomenon (Heckman et al.
1990).

On one hand, this interpretation raises a basic question: what is the
mechanism producing such high plasma temperatures ($kT \sim 6$ keV in NGC
253 and $\sim$ 8 keV in M82) and high luminosities? Indeed, hydrodynamical
simulations of superwinds do predict HEX emission at such high
temperatures, but with a lower than observed HEX/LEX luminosity ratio
(Suchkov et al. 1994).

On the other hand, this interpretation is the most straightforward and can
explain {\it all} the observational results. The chemical abundances of
the HEX component, consistent with those deduced for the LEX component,
are likewise in agreement with the trend expected for SB-driven superwind
models (see Sect. 8.1) and hence are compatible with a SN origin for (a
substantial fraction of) the HEX component.  Of course, if the HEX
component of thermal X-ray emission is found to originate {\it in} the
wind material ejected from the starburst (rather than from the
wind-shocked interstellar material), one may expect a higher abundance of
heavy elements (Suchkov et al. 1994). In both galaxies, however, Fe in the
nuclear regions could have been strongly depleted (relative to lighter
elements) by the plentiful dust and warm ISM clouds (Telesco 1988, P97).  
The superwind model also offers a natural explanation for both the
$r^{-2}$ X-ray brightness profile (Sect. 4) and temperature profile (Sect.
6)  of NGC 253 since it predicts that the pressures/densities should
decline systematically at large radii, thus reducing both emissivity and
temperature, as observed (Heckman et al. 1990).

The substantial photoelectric absorption might seem difficult to explain
in the (extended) superwind scenario. However, one should consider the
following: {\it i)} given our limited spatial resolution, it's still
possible that absorption may affect only the fraction of the HEX component
that comes from the nuclear region; {\it ii)} similar results are found
for the GRXE (see Kaneda 1998 for estimates on how such absorption is
distributed along the Galactic plane and largely affects also the outer
parts of the Galactic disk); and {\it iii)} there are several lines of
evidence for molecular gas/dust outflows reaching out several kpc from the
nucleus (Alton et al. 1999; Wei$\ss$ et al. 1999). Independent estimates
of the X-ray absorption column densities can be obtained from the
extinction measured in (e.g.) the K band (2.2 $\mu m$ peak), using the
conversion formula: $N_H/A_K$ = $1.8\times 10^{22}$ cm$^{-2}$ (Telesco et
al. 1991 and ref. therein). Telesco et al. (1991) and Rieke et al. (1980)
applied such formula to M82 and NGC 253 obtaining N$_H$ $\sim$ $0.7\times
10^{22}$ cm$^{-2}$ and $1.8\times 10^{22}$ cm$^{-2}$ respectively,
consistent with our derived values.

In summary, although it is not clear yet whether a superwind scenario could
explain the origin of the observed HEX emission, the simplicity of this
interpretation, along with the analogy with the GRXE, make this model
somewhat more attractive. In the following, therefore, we explore the
implications of a superwind scenario as the main origin of the HEX
emission.

\subsubsection{Estimates and implications from the Superwind model}

If one assumes that a superwind-driven hot, diffuse, gas is responsible
for the bulk of the HEX emission, the first immediate consequence is that
such thermal gas cannot be confined by gravity. Following Wang et al.
(1995), the observed temperatures ($T_{\rm obs}$ $\sim$ 6.7/9.5 $\times$
10$^7$ K for NGC253/M82) are in fact much higher than the ``escape
temperature'' ($T_{\rm esc}$ $\sim$ 2/1 $\times$ 10$^6$ K for NGC253/M82;
Wang et al. 1995) of the gas in these galaxies.

>From the normalizations of the HEX components (col. 2 of Table 4), one can
derive some other interesting parameters (see Table 4): the gas average
density ($n_e$) and pressure ($p$), total mass ($M_{\rm tot}$), radiative
cooling time ($t_{\rm cool}$), and bulk velocity ($v_{\rm bulk}$). For these
calculations, we used the simplified formulae from Della Ceca et al. (1997)
except for $t_{\rm cool}$ that we take from equation 5.23 of Sarazin (1988).
[Average values have been computed over the X-ray emitting region assuming a
sphere with radius $r = 3$ kpc in both galaxies (corresponding to $\sim$ 1.5
$r_c$, see Sect. 4), and parameterizing the gas clumpiness by a filling
factor $f \equiv {\rm volume~ of~ X-ray~ emitting~ gas\over{\rm volume~ of~
X-ray~ emitting~ region}}<1$. We also provide in Table 4 the mass outflow
rate ({\it \.M$_{\rm bulk}$})  from $$ \dot M_{\rm bulk} \sim 4 \pi r^{2}
\cdot v_{\rm bulk} \cdot n_{e,3kpc} \cdot m_{p} \cdot f^{-1/2},$$ where the
gas density at 3 kpc $n_{e,3kpc}$ ($\sim {1\over{2.7}} n_e$) has been
calculated assuming a density profile of the form $$n_e \propto
[1+({r\over{r_c}})^2]^{-1}$$ (see Sect. 4 and equation 5.63 in Sarazin 1988).

We note that both the Fe and $\alpha$-element abundances are similar to those
found in cluster of galaxies.  From the gas bulk velocity and cooling time,
one can derive a typical dimension $v_{\rm bulk}$ $\cdot$ $t_{\rm cool}$
$\sim$ 3 Mpc, where the gas expelled from the galaxies can be redistributed.
This value is of the order of the size of cluster of galaxies suggesting the
possible relevance of this observation for the problem of the ICM metal
enrichment.

However, assuming a starburst lifetime of $\sim$ 10$^7$ yr for both 
NGC 253 (Engelbracht et al. 1998) and M82 (Satyapal et al. 1997), and
{\it \.M$_{\rm bulk}$} $\sim$ 24 (52)  M$_{\odot}$ yr$^{-1}$ for NGC 253
(M82), the mass ejected would be $\sim$ 2.4 (5.2) $\cdot$ 10$^8$
M$_{\odot}$.  Taken at face value, such a ``small'' ejected mass would
eventually require too large a number ($\gsimeq$ 10$^5$) of SBGs to make
all of the $\sim$ 10$^{14}$ M$_{\odot}$ of a typical cluster, but may
nevertheless be significant.  Given the $\sim$ 0.1 yr$^{-1}$ SN rates of
NGC 253 and M82, and assuming that each SN ejects $\sim$ 10 M$_{\odot}$ of
gas, most of the gas ($>$ 95\%) expelled by the galaxies must necessarily
be ambient ISM gas ``swept-up'' by the superwind.  The low Fe abundances
derived here may therefore be indicative of low abundances in the disk ISM
involved in the starburst.

In the following, we check whether the SNe-driven superwind scenario is
energetically self-consistent.  The total thermal energy required to heat the
gas of NGC 253 (M82)  is $E_{\rm th} \sim kT \cdot n_e \cdot {4\over{3}} \pi
r^3$ $\sim$ 6 (16)  $\times$ 10$^{56}$ erg. These numbers are of the order of
the total energy available from SNe ($E_{\rm th, SN} \sim 10^{57}$ erg),
assuming a SN rate of $\sim$ 0.1 SN yr$^{-1}$, that each SN releases $\sim$
10$^{51}$ erg, and a typical burst duration of 10$^7$ yr, and indicate that
SNe {\it alone} may have supplied the total energy present today in the
superwind. This agreement holds, however, if one assumes the gas to be in
equilibrium and stationary while, as outlined above, the high gas temperature
implies that the gas is probably being expelled out of the galaxy. One can
then approximate the thermal energy lost in the outflow \.E$_{\rm th}$ as
{\it \.E$_{\rm k} \simeq kT \cdot {\dot M_{bulk} \over{m_{\rm p}}}$} $\simeq$
3 (7) $\cdot$ 10$^{50}$ $f^{1/2}$ erg yr$^{-1}$ for NGC 253 (M82). This value
should be compared with the maximum energy rate of $\sim$ 10$^{50}$ erg
yr$^{-1}$ available from SNe. Therefore, the superwind hypothesis faces a
slight energy replenishment problem. It should be noted that the same problem
is not present for the LEX component since in this case, both temperature
and mass outflow rate are lower by about an order of magnitude, thus
resulting in a thermal energy loss rate $\sim$ 100 times lower which is
thereby more than consistent with the energy available from SNe. A solution
to this problem may come from one of the following possibilities: {\it i)}
additional energy sources may come from either the intense stellar winds of
``pre-SN stars'', and/or intense UV flux by young stars, and/or a higher SN
rate in the past; {\it ii)} the filling factor is $\ll$ 1;  {\it iii)} the
gas does {\it not} escape from the galaxy but is confined to it 
(e.g. by some higher than usual interstellar clouds/gas pressure); or {\it
iv)} there is a more exotic powering mechanism, as outlined below. In
conclusion, our simplified estimates indicate no strong inconsistency in the
starburst-driven superwind scenario.

Finally, we wish to point out also the similarities of the present results
with the X-ray luminous SBGs NGC 3310 and NGC 3690 that exhibit very similar
X-ray spectra (Zezas et al. 1998). As a matter of fact, the HEX components
found in these sources are statistically (although not significantly) better
fit with a thermal model (with $kT_{\rm 2-10 keV} \sim$ 10 keV) rather than
with a power-law model.  This suggests that the above considerations may
apply also to higher luminosity SBGs and, as such, may be rather general to
SBGs.

\subsubsection{Other mechanisms}

Based on recent {\it RXTE} data on the non-thermal component in the
GRXE HEX flux, which contributes $\sim$50\% of the total
3--10 keV spectrum and reaches up to $\sim$30 keV, Valinia \& Marshall
(1998) propose an explanation in terms of non-thermal bremsstrahlung
emission by energetic electrons accelerated by SNRs. However, this would
still require a ``spectral conspiracy'' as discussed in Sect. 8.2.5, and
is furthermore constrained by the PDS upper limits above 10 keV. More
appealing could be another mechanism, also first proposed to explain the
GRXE observations (Tanaka et al. 1999), that could explain the HEX
continuum plus 6.7 keV line: charge exchange by low-energy heavy ions with
neutral gas in the galaxies. A magnetic confinement picture has been
proposed by Makishima (1994, 1996) to explain, again, the production of
the GRXE HEX emission. In this model, some fraction of the cooler
plasma, confined by magnetic loops in the galactic disk, is heated to the
observed temperature by the dissipation of the galaxy's rotational and
velocity dispersion energy through magnetic compression and reconnection
(see Tanuma et al. 1999). This solution has the merit of providing a
natural explanation for the confinement of the gas (through the magnetic
field) and its heating to high temperatures.

More quantitative work on these models are, however, required in order to
estimate their applicability also to SBGs.

\section{Summary}

NGC 253 and M82 have been observed by {\it BeppoSAX} in the 0.1--60 keV
energy range. The spectra are complex, exhibiting several emission lines,
and including a LEX component and an absorbed HEX one. The analysis of the
LEX emission is problematic (due to limited statistics and low spatial
resolution), but is consistent with emission from SNRs, X-ray binaries and
mildly-hot diffuse gas heated by a SN-driven superwind, as found by other
authors (e.g. P97 and ref. therein). As for the latter, the detection of
ionized Fe-K emission lines and high-energy continuum cut-offs in the
spectra of both SBGs, and the evidence that such HEX component is
extended, indicate strongly that a thermal hot ($kT$ $\sim$ 5--8 keV)
plasma produces most of the HEX emission ($L_{2-10 keV}$ $\sim$ 1--5
$\times$ 10$^{40}$ ergs$^{-1}$).  Fe abundances are found to be $\sim$ 0.3
solar in NGC 253 and $\sim$ 0.07 solar in M82. A contribution by an
additional non-thermal absorbed power-law component is possible, but the
present analysis suggests that this is $\lsimeq$20\%.

We have also discussed several possible alternative explanations for the
origin of the HEX component. Possible candidates include SNRs, Compton
scattering, XRBs, absorbed LLAGN, and superwind thermal emission to
varying degrees.  More exotic explanations such as charge exchange
processes, non-thermal bremsstrahlung emission or magnetic reconnections
have also been discussed. We have shown however that, in spite of some
inconsistencies with hydrodynamical models predictions, direct thermal
emission from a hot superwind is a likely candidate for the origin of the
HEX component. It has the advantage that it can explain most naturally and
directly the observational results and does not require any spectral
fine-tuning to produce the observed thermal 3--10 keV spectrum.  We
evaluated the energetic balance taking into account the various energy
sources available during the starburst (e.g. SNe explosions, stellar winds
and intense UV flux from young stars) and concluded that there is
currently no severe inconsistency with the proposed starburst-driven
superwind model.

Altogether, our results highlight the significance of high-energy
phenomena in actively star-forming galaxies, as previously predicted (e.g.
Bookbinder et al. 1980; Heckman et al. 1990, 1993 and ref. therein).

\begin{acknowledgements}
We acknowledge partial support by the Italian Space Agency under the contract 
ASI-ARS-98-119 and by the Italian Ministry for University and Research (MURST) 
under grant Cofin-98-02-32We acknowledge financial support from ASI and MURST.
MC thanks R. DellaCeca, F. Fiore, L. Ciotti, A. D'Ercole and
S. Mariani, for helpful discussions.
\end{acknowledgements}

%\end{document}

\onecolumn

\begin{table}[htb]
\begin{center}
{\bf Table 2:} Basic Line Parameters$^a$ \\
\vspace{0.2cm}
\begin{tabular}{cc|cc|c}
\hline
\hline
\multicolumn{2}{c|}{NGC 253} &
\multicolumn{2}{|c}{M82} &
\multicolumn{1}{|c}{} \\
\hline
Obs. E & Intensity & Obs. E & Intensity & Possible Identifications \\
(keV) & ($\times$10$^{-4}$ ph. cm$^{-2}$s$^{-1}$) & (keV) & ($\times$10$^{-4}$ ph. cm$^{-2}$s$^{-1}$) & 
(nearest K$_{\alpha}$ lines of H- and He-like ions)\\
\hline
\multicolumn{4}{c}{\underline{Low Energy Lines$^b$}} &
\multicolumn{1}{|c}{} \\
& & 0.51$^{+0.18}_{-0.17}$ & 4.4$^{+2.6}_{-2.4}$ & O{\scriptsize VII} (0.57 keV)\\
0.66$^{+0.03}_{-0.03}$ & 3.9$^{+1.3}_{-1.7}$ & & & O{\scriptsize VIII}(0.65 keV)\\
& & 0.72$^{+0.02}_{-0.02}$ & 11.3$^{+2.2}_{-1.7}$ & 
O{\scriptsize VIII},Ne{\scriptsize IX} and less ionized species \\
0.90$^{+0.03}_{-0.05}$ & 2.7$^{+0.7}_{-1.0}$ & 0.89$^{+0.01}_{-0.02}$ & 15.5$^{+1.6}_{-2.6}$ & 
Ne{\scriptsize IX}(0.918 keV)\\
1.10$^{+0.08}_{-0.09}$ & 0.8$^{+0.4}_{-0.5}$ & 1.09$^{+0.02}_{-0.03}$ & 5.9$^{+0.7}_{-1.1}$ & 
Ne{\scriptsize X}(1.02 keV)-FeL ($\sim$ 0.8-1.1 keV)\\
1.39$^{+0.09}_{-0.08}$ & 0.4$^{+0.2}_{-0.3}$ & 1.35$^{+0.03}_{-0.04}$ & 2.9$^{+0.5}_{-1.2}$ & 
Mg{\scriptsize XI}(1.35keV)-Mg{\scriptsize XII}(1.47 keV)\\
1.88$^{+0.04}_{-0.03}$ & 0.3$^{+0.2}_{-0.1}$ & 1.88$^{+0.03}_{-0.01}$ & 2.2$^{+0.2}_{-0.4}$ & 
Si{\scriptsize XIII}(1.86 keV)\\
2.41$^{+0.06}_{-0.06}$ & 0.2$^{+0.1}_{-0.1}$ & 2.49$^{+0.06}_{-0.06}$ & 0.8$^{+0.1}_{-0.2}$ & 
S{\scriptsize XV}(2.45 keV)\\
& &  3.65$^{+0.15}_{-0.17}$ & 0.2$^{+0.2}_{-0.1}$ & Ar{\scriptsize XVI}(3.13 keV)-Ca{\scriptsize IXX}
(3.85 keV)\\
\multicolumn{4}{c}{\underline{Fe-K$_{\alpha}$  Lines$^c$}} &
\multicolumn{1}{|c}{} \\
%E (keV) & EW (eV) & ID & E (keV) & EW (eV) & ID\\
6.4 (fixed) & $<$0.04 &  6.4 (fixed) & $<$ 0.15 & Fe I (6.4 keV)\\
6.69$^{+0.07}_{-0.06}$ & 0.16$^{+0.03}_{-0.03}$ & 6.63$^{+0.21}_{-0.20}$ & 0.16$^{+0.11}_{-0.10}$ & 
Fe XXV (6.69 keV)\\
\hline
\hline
\end{tabular}
\end{center}
\end{table}
Note: Intervals are at 90\% confidence for one interesting
parameter.

$^a$ Obtained by fitting the MECS+LECS data with a continuum model consisting in a 
two-temperature bremsstrahlung model with temperature and absorption fixed to the 
best-fit values obtained in Sect. 5 (see Table 3).

$^b$ Equivalent widths were ranging from a few hundred eV to a few tens eV. 

$^c$ Equivalent widths of the detected Fe-K lines at $\sim$ 6.7 keV were 
400$^{+125}_{-75}$ eV for NGC 253 and 60$^{+38}_{-45}$ eV for M82.

\begin{table}[htb]
\begin{center}
{\bf Table 3:} Best-fit two-components (vmekal) thermal models \\
\vspace{-0.2cm}
\begin{tabular}{ccccccccc}
&&&& \\
\hline
Source & $kT$$_{soft}$ & Ab$_{soft}$ &  $N_{\rm H}^{\rm hard}$ & $kT$$_{hard}$ & 
Ab$_{hard}^a$ & $\chi^2/dof$ & $F_{\rm 0.1-2 keV}$/$F_{\rm 2-10 keV}$ & $L_{\rm 0.1-2 keV}$/$L_{\rm 2-10 keV}$ \\
 & keV & $\alpha-$el. &  $\times 10^{22}$ & keV & $\alpha-$el. & & $\times$10$^{-12}$ & $\times$10$^{40}$ \\
 & &  Fe, Ni & cm$^{-2}$ & & Fe, Ni & & erg cm$^{-2}$s$^{-1}$ & erg s$^{-1}$ \\ 
\hline 
NGC253 & $0.81_{-0.07}^{+0.13}$ & $1.57_{-0.71}^{+0.93}$ & $1.19_{-0.39}^{+0.32}$ & 
$5.75_{-0.48}^{+0.56}$ & $1.57_{-0.71}^{+0.93}$ & 191/180 & 3/4.8 & 0.8/1.5 \\
%&&&&&& &  & & & \\
& & $0.17_{-0.09}^{+0.10}$ & & & $0.32_{-0.10}^{+0.09}$ & &  &\\
&&&&&& &  & \\
 M82 & $0.70_{-0.05}^{+0.05}$ & $2.89_{-1.00}^{+1.44}$ & $0.58_{-0.15}^{+0.14}$ 
& $8.20_{-0.58}^{+0.59}$ & $2.89_{-1.00}^{+1.44}$ & 169/135 & 11.4/29 & 1.6/5.9\\
%&&&&&& & & & & & \\
& & $0.85_{-0.28}^{+0.50}$ & & & $0.07_{-0.03}^{+0.05}$ & & &\\
%&&&&&& & &\\
\hline
\end{tabular}
\end{center}
%Note: The value of the relative normalizations $A_{LECS}/A_{MECS}$ and 
%$A_{MECS}$/$A_{PDS}$ are $\simeq$ $0.66$ $\pm$ 0.03 and $\simeq$ ??, 
%consistent with the prescriptions with the Sax Data Center (Fiore, Grandi 
%\& Guainazzi 1998) 
\end{table}
Note: Intervals are at 90\% confidence for one interesting
parameter.

$^a$ Ab$_{hard}$ was set equal to Ab$_{soft}$ for the $\alpha$
elements.

\begin{table}[htb]
\begin{center}
{\bf Table 4:} Physical conditions of the hot X-ray emitting gas$^a$ \\
\vspace{0.2cm}
\begin{tabular}{cccccccc}
\hline
Source 	& K$^b$ & $n_e$ & $p$ & $M_{tot}$ & $t_{\rm cool}$ & $v_{\rm bulk}$ & {\it \.M$_{\rm bulk}$} \\
	& ($\times 10^{-3}$) & ($\times f^{-1/2}$ cm$^{-3}$) & ($\times f^{-1/2}$ dyne cm$^{-2}$) & ($\times f^{1/2}$ M$_{\odot}$) & 
($\times f^{-1/2}$ yr) & ($\times$ km s$^{-1}$) & ($\times f^{-1/2}$ M$_{\odot}$ yr$^{-1}$)\\
\hline
NGC 253 & 4.5 & 2.2$\times 10^{-2}$ & 4$\times 10^{-10}$ & 4.6 $\times 10^{7}$ & 3.2$\times 10^{9}$ & 1050 & 24 \\
M82     & 22  & 3.9$\times 10^{-2}$ & 1$\times 10^{-9}$ & 8.3 $\times 10^{7}$ & 2.1$\times 10^{9}$ & 1250 & 52 \\
\hline
\end{tabular} 
\end{center}
\end{table}

$^a$ The values are averaged over the X-ray emitting volume (sphere with 
radius $r=3$ kpc)

$^b$ normalizations of the hard components at 1 keV, 
in units of [$10^{-14}/(4\pi D^2)] \cdot \int n_{e}^{2} dV$, where $D$ is 
the distance to the source in cm, $n_e$ is 
the electron density in units of cm$^{-3}$, and $V$ is the volume filled 
by the X-ray emitting gas in cm$^{-3}$.

%{\it \.M$_{\rm bulk}$}, the mass bulk outflow rate, has been estimated using 
%the simplified formula: {\it \.M$_{\rm bulk}$} $\sim 4 \pi r^{2} v_{\rm bulk} 
%\Delta t m_{p} f^{-1/2}$ 
%M$_{\odot}$ yr$^{-1}$, where $r$ is the source radius and $\Delta t$ is 1 yr.

\begin{figure}[htb]
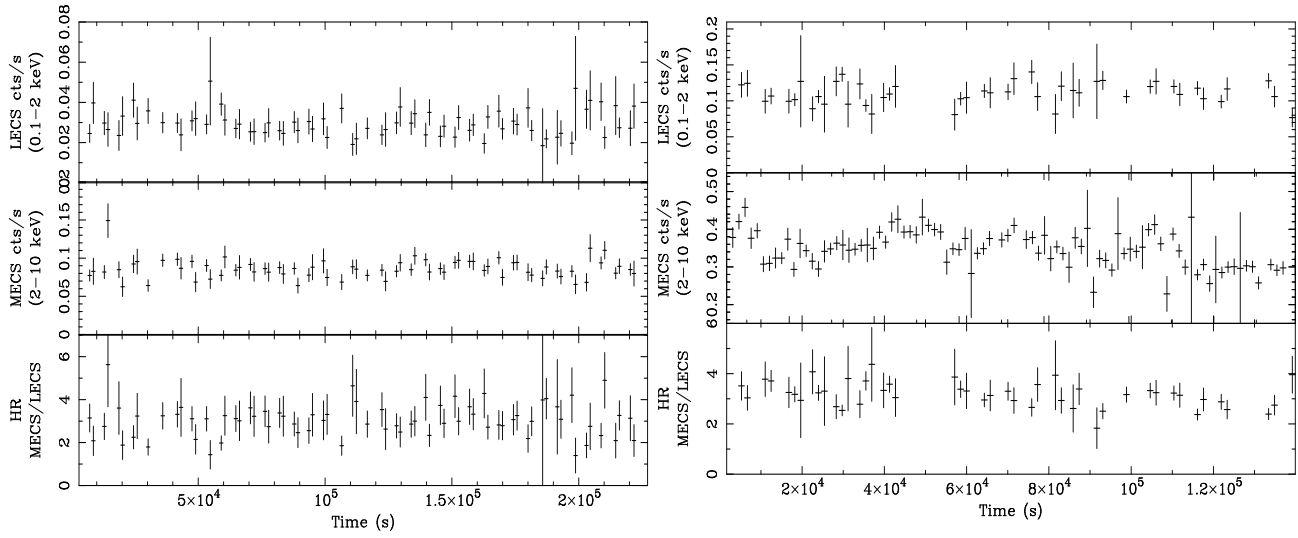

\parbox{9truecm}
{\psfig{file=./ngc253_lecs_mecs_hr.ps,width=8.5cm,height=7cm,angle=-90}}
\parbox{9truecm}
{\psfig{file=./m82_lecs_mecs_r30_hr.ps,width=8.5cm,height=7cm,angle=-90}}
\caption[h]{Light curves of NGC 253 (left) and M82 (right).}
\end{figure}

\begin{figure}[htb]
\parbox{9truecm}
{\psfig{file=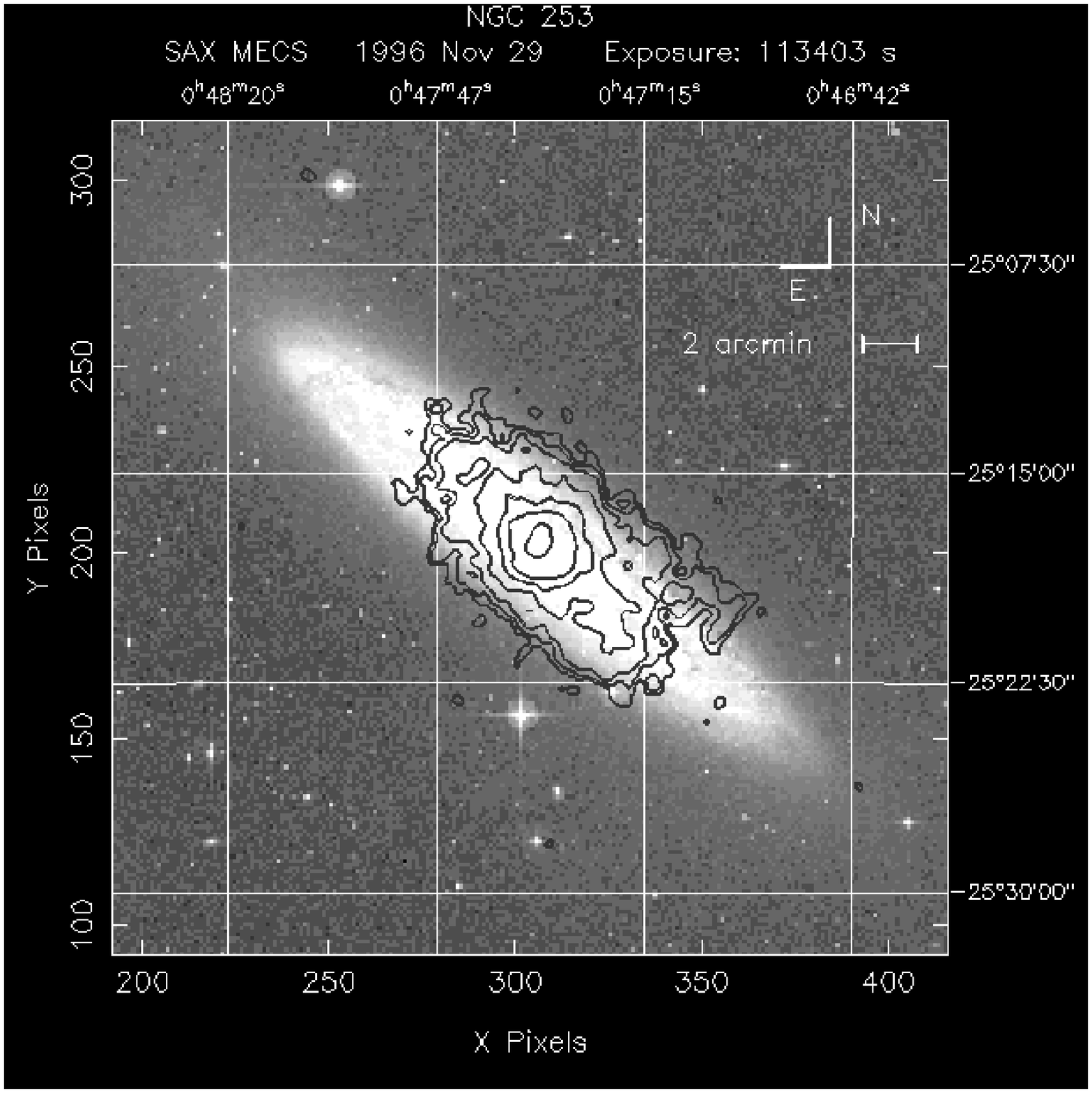,width=8.5cm,height=8.5cm,angle=0}}
\parbox{9truecm}
{\psfig{file=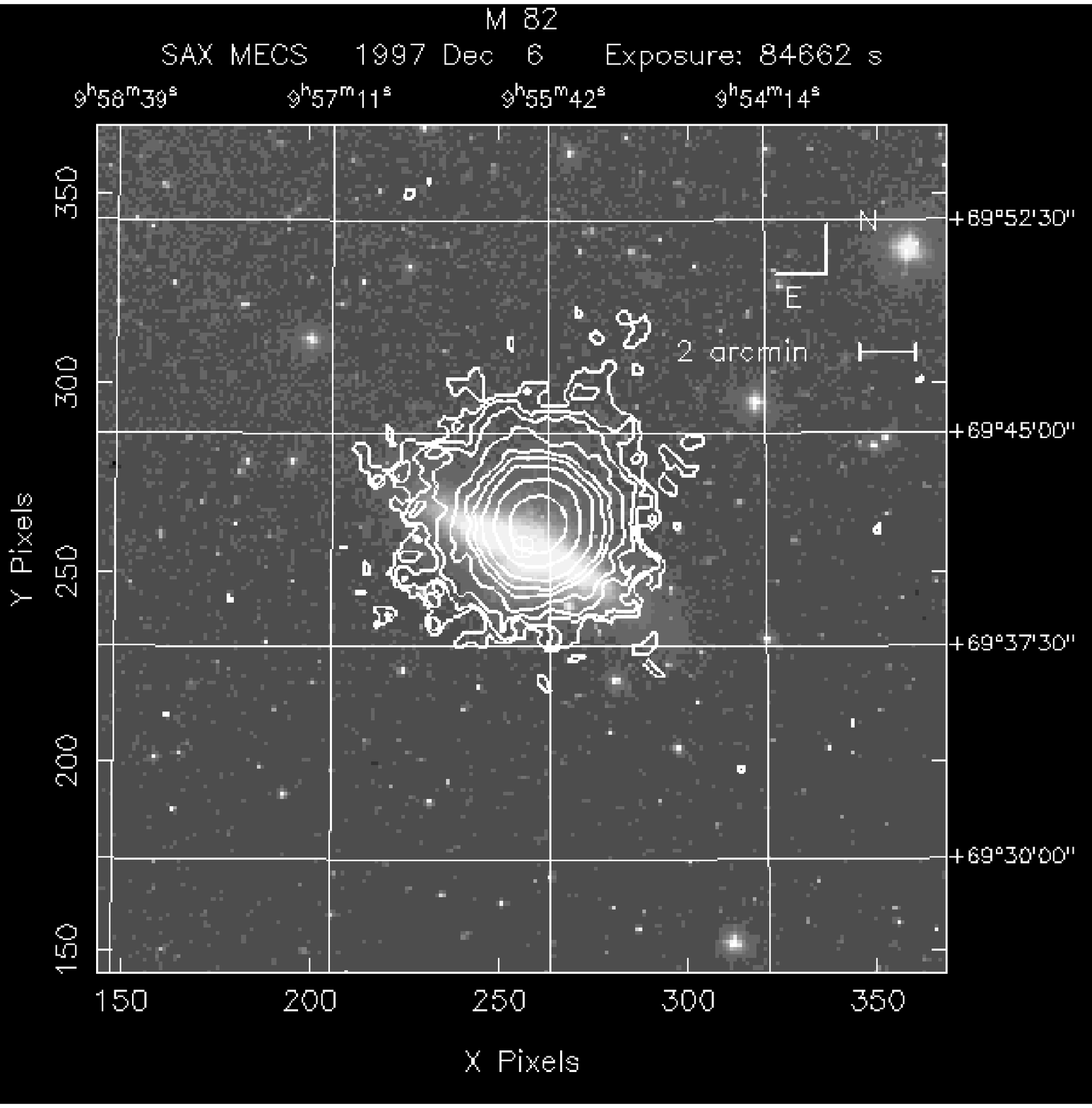,width=8.5cm,height=8.5cm,angle=0}}
\caption[h]{{\it BeppoSAX} MECS 3--10 keV images of NGC 253 (left) and M82 (right), 
superimposed on optical DSS images. Both FOVs are $\sim$ 30$^{\prime}$ 
$\times$ 30$^{\prime}$, and correspond to $\sim$ 22 $\times$ 22 kpc for NGC 253 and 
26 $\times$ 26 kpc for M82. Plotted are the 2$\sigma$, 3$\sigma$, 5$\sigma$, 
10$\sigma$, 20$\sigma$, 30$\sigma$, 50$\sigma$ contours for NGC 253, and 
the 2$\sigma$, 3$\sigma$, 5$\sigma$, 10$\sigma$, 20$\sigma$, 30$\sigma$, 50$\sigma$, 
100$\sigma$, 200$\sigma$ contours for M82.
The apparent shift between the centroids of the X-ray contours 
and the DSS image of M82 is within the systematics (of $\sim$ 1$^{\prime}$) in the 
absolute position determination for {\it BeppoSAX}.}
\end{figure}

\begin{figure}[htb]
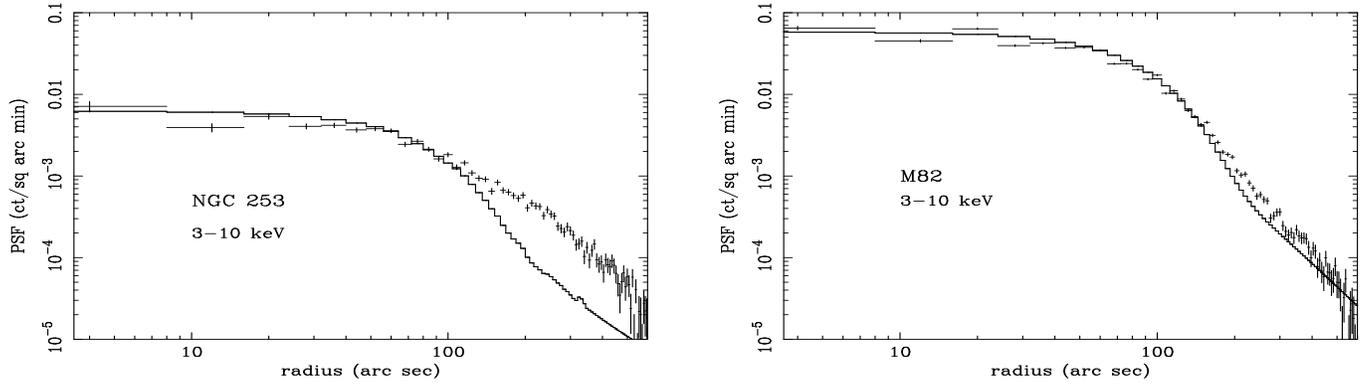

\vspace{1truecm}
\parbox{8truecm}
{\psfig{file=./ngc253_rad_prof_3_10.ps,width=8.5cm,height=5cm,angle=-90}}
\ \hspace{0.5truecm} \
\parbox{8truecm}
{\psfig{file=./m82_rad_prof_3_10.ps,width=8.5cm,height=5cm,angle=-90}}
\caption[h] {Radial profile (data points) of the source emission in the 3--10 keV band 
compared with the PSF for an on-axis point source (solid line) calculated integrating 
over 3--10 keV but weighting the different energies with the sources spectra. 
Extended emission is clearly detected in NGC 253 but only marginally in M82.}
\end{figure}

\begin{figure}
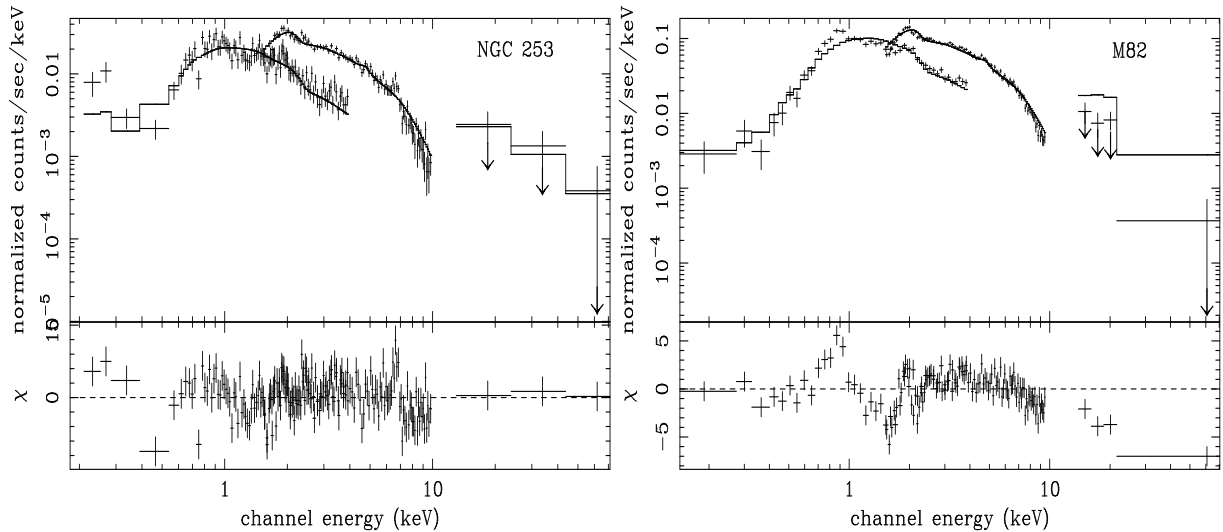

\parbox{9truecm}
{\psfig{file=./ngc253_pldat_po.ps,width=8cm,height=7cm,angle=-90}}
\parbox{9truecm}
{\psfig{file=./m82_pldat_po.ps,width=8cm,height=7cm,angle=-90}}
\caption[h]{The spectra of NGC 253 and M82 fitted with a single power-law model ($\Gamma$ 
$\sim$ 1.7) to highlight the spectral complexities (emission lines and curvature of the 
continuum). All the data, including the PDS data, are binned such as to have a 
{\it statistical} significance $>$ 2$\sigma$ per bin. If systematic errors are included, 
the PDS data provide only upper-limits (arrows).}
\end{figure}

\begin{figure}
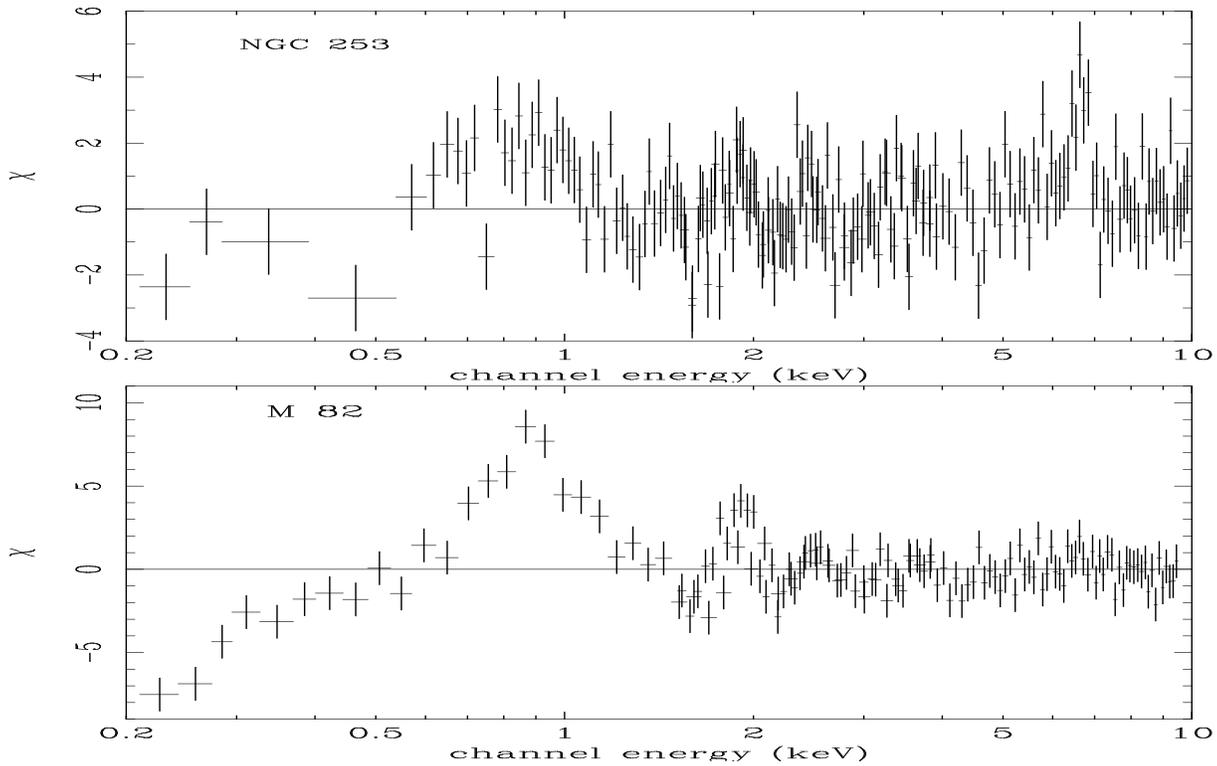

{\psfig{file=./ngc253_2tbrem_res.ps,width=16cm,height=5cm,angle=-90}}
{\psfig{file=./m82_2tbrem_res.ps,width=16cm,height=5cm,angle=-90}}
\caption[h]{Residuals obtained by fitting NGC 253 (top) and M82 (bottom) 
spectra with a 2-T bremsstrahlung continuum model, with $kT$$_{soft}$, $kT$$_{hard}$ and 
N$_{H}$ equal to the best-fit values shown in Table 3. Strong analogies between 
the two spectra are quite apparent.}
\end{figure}

\begin{figure}
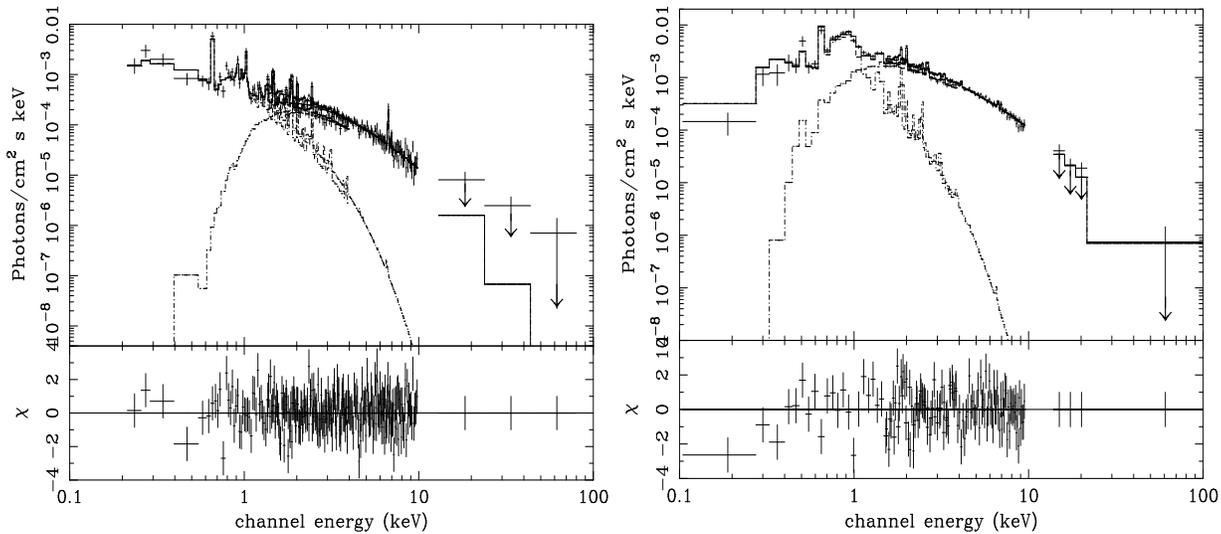

\parbox{9truecm}
{\psfig{file=./cappi_fig2a.ps,width=8cm,height=7cm,angle=-90}}
\parbox{9truecm}
{\psfig{file=./m82_2tbrem_ufspec_del.ps,width=8cm,height=7cm,angle=-90}}
\caption[h]{Best-fit unfolded spectra of NGC 253 (left) and M 82 (right) obtained with a 
2-T Mekal model (Table 2). The data are binned such as to have a S/N $>$ 2 in each bin. 
Only data below 10 keV are used in the fitting, PDS data being 
added subsequently.}
\end{figure}

\begin{figure}
\psfig{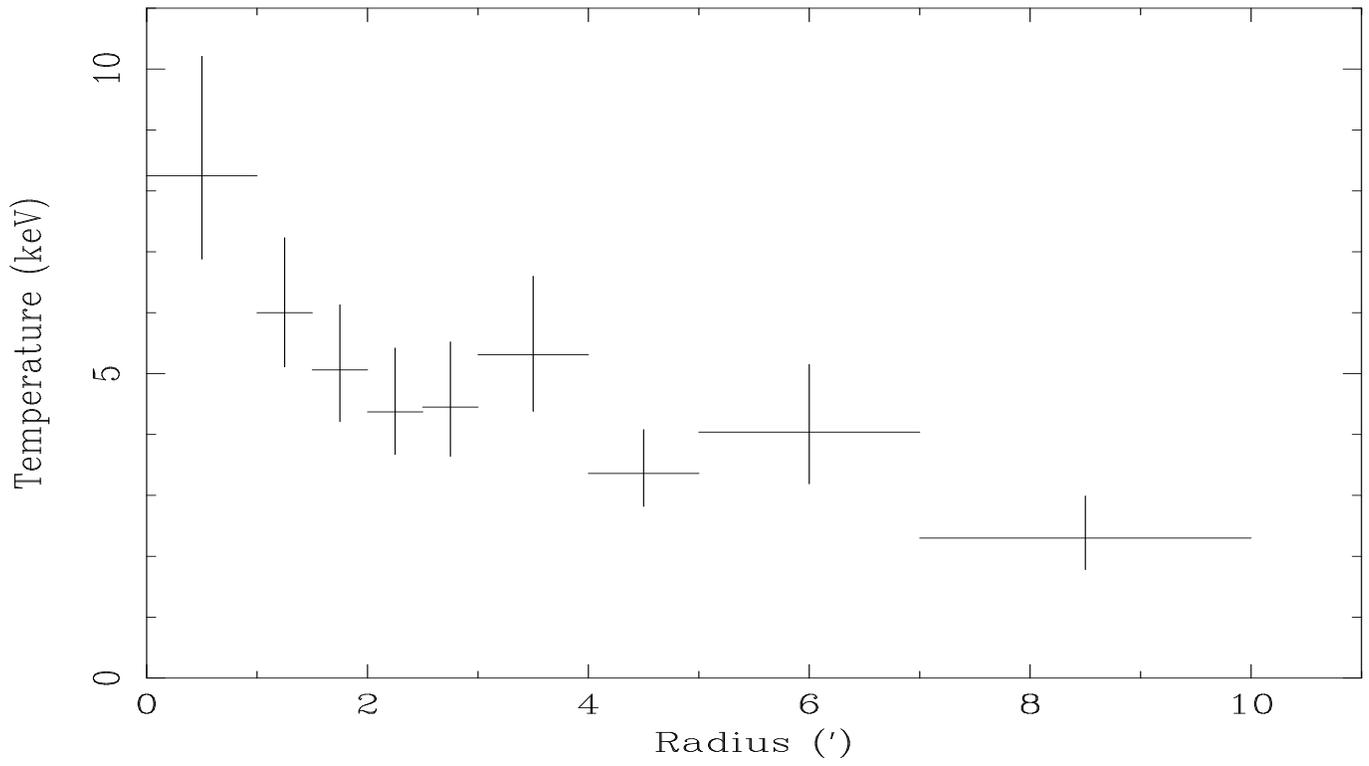}
\caption[h]{Temperature profile of NGC 253}
\end{figure}

\end{document}